\documentclass[aps,pra,twocolumn,longbibliography]{revtex4-1}

\usepackage{amsmath}
\usepackage{mathtools}
\usepackage{braket}
\usepackage{bbm}
\usepackage{graphicx}
\usepackage{amssymb}
\usepackage[dvipsnames]{xcolor}
\usepackage{multirow}
\usepackage{xfrac}
\usepackage{textcomp}
\usepackage{comment}

\usepackage{hyperref}
\usepackage{bbold}

\hypersetup{
	colorlinks=true,
	linkcolor=blue,
	citecolor=blue,
	filecolor=magenta,
	urlcolor=cyan,
	breaklinks=true
}
\newcommand{\papertitle}{Topological superfluid transition in bubble-trapped condensates}
\newcommand{\unipd}{Dipartimento di Fisica e Astronomia ``Galileo Galilei'', Universit\`a di Padova, via Marzolo 8, 35131 Padova, Italy}
\newcommand{\tuk}{Physics Department and Research Center OPTIMAS, Technische Universit\"at Kaiserslautern, Erwin-Schr\"odinger Stra\ss e 46, 67663 Kaiserslautern, Germany}
\newcommand{\inocnr}{Istituto Nazionale di Ottica (INO) del Consiglio Nazionale delle Ricerche (CNR), \\ via Nello Carrara 1, 50125 Sesto Fiorentino, Italy}
\newcommand{\infn}{Istituto Nazionale di Fisica Nucleare (INFN), Sezione di Padova, 
via Marzolo 8, 35131 Padova, Italy}
\newcommand{\lptms}{Universit\'e Paris-Saclay, CNRS, LPTMS, 91405 Orsay, France}

\DeclareSymbolFont{sfletters}{OML}{cmbrm}{m}{it}

\DeclareMathSymbol{\matrrho}{\mathord}{sfletters}{"1A}

\begin{document}
	
\title{\papertitle}
\author{Andrea Tononi}
\affiliation{\unipd}
\affiliation{\infn}	
\affiliation{\lptms}
\author{Axel Pelster}
\affiliation{\tuk}
\author{Luca Salasnich}
\affiliation{\unipd}
\affiliation{\inocnr}
\affiliation{\infn}

\date{\today}

\begin{abstract}
Ultracold quantum gases are highly controllable and, thus, capable of
simulating difficult quantum many-body problems ranging from condensed
matter physics to astrophysics. Although experimental realizations have so
far been restricted to flat geometries, recently also curved quantum
systems, with the prospect of exploring tunable geometries, are produced in
microgravity facilities as ground-based experiments are technically
limited. Here we analyze bubble-trapped condensates, in which the atoms
are confined on the surface of a thin spherically-symmetric shell by means
of external magnetic fields. A thermally-induced proliferation of
vorticity yields a vanishing of superfluidity. We describe the occurrence
of this topological transition by conceptually extending the theory of
Berezinskii, Kosterlitz and Thouless for infinite uniform systems to such
finite-size systems. Unexpectedly, we find universal scaling relations for
the mean critical temperature and the finite width of the superfluid
transition. 
Furthermore, we elucidate how they could be experimentally
observed in finite-temperature hydrodynamic excitations.
\end{abstract}

\maketitle 

\section{Introduction}
The physical understanding of Nature, since the birth of modern science, relies on the reduction of a complex system to a properly idealized model that allows a mathematical description. 
In principle, it is possible to engineer different physical systems in such a way that they are described by the same laws: the study of each system, therefore, will offer important information on the others. 
These basic concepts are at the core of Feynman's visionary idea \cite{feynman} of simulating the complex behavior of many-body systems with controllable algorithms that, rather than being implemented in classical computers, are based on the laws of quantum mechanics \cite{georgescu,lloyd}. 
Nowadays, the experimental techniques in the field of ultracold atomic gases allow to tune and control every term of the Hamiltonian \cite{bloch},  either kinetic, potential or interaction ones. 
On one hand, it is thus possible to explore the different physical regimes of quantum many-body systems by engineering the Hamiltonian with external electric and magnetic fields  \cite{greiner,bloch2}. 
On the other, the Hamiltonian of experimentally-inaccessible systems can be mapped to analogue models for ultracold atoms \cite{liberati}: the Hawking radiation \cite{hawking1}, for instance, was observed for the first time in analogue black holes \cite{steinhauer,denova}. 

Until now, most experiments with ultracold quantum gases focused on the investigation of flat geometries. The prospect of producing two-dimensional superfluid manifolds \cite{moller}, whose geometry and curvature can be properly tuned, would offer, besides the intrinsic scientific interest, additional degrees of freedom for simulating physical systems. 
Considering two-dimensional curved gases in regimes of quantum degeneracy, it is quite natural to ask if the delicate interplay of curvature, geometry and interactions allows for the superfluid properties to emerge \cite{machta1989}. 
In particular, it is unclear whether and how the Berezinskii-Kosterlitz-Thouless transition \cite{berezinskii,kosterlitz1973,kosterlitz1974} (BKT) of the superfluid density occurs in curved closed shells, and if its driving mechanism is the thermally-driven unbinding of vortex-antivortex excitations also for these finite-size systems. 
This peculiar topological transition, even for infinite-system sizes, does not affect the thermodynamic functions, but results in a universal jump of the superfluid density at a critical temperature \cite{nelson}. 
Describing the BKT transition and its main experimental consequences in the simple paradigmatic case of a curved spherical shell, thus, would set a milestone in the development of quantum many-body physics with ultracold atoms. 

Here we focus our theoretical investigation on the quantum statistical and topological properties of bubble-trapped Bose-Einstein condensates \cite{zobay}, which result from cooling and confining atomic gases on thin two-dimensional shells \cite{lundblad,carollo2021,guo2021}. 
The experimental study of these systems, which are technically difficult to produce in the presence of gravity \cite{colombe,demarco}, is made possible by the recent development of microgravity settings for studying Bose-Einstein condensates, either space-based \cite{elliott,aveline,becker} as the Cold Atom Lab (CAL) on the International Space Station, or {ground-based free-falling ones} \cite{vanzoest,elevator}.
Around the temperature of the topological BKT transition we find here the
emergence of universal laws in microgravity bubble-trapped superfluids,
despite the naive expectation of nonuniversal, i.~e.~system-dependent,
physics.
For different shell sizes and densities we predict a data collapse for
both the critical temperature and the transition width. This
theoretical finding is immediately testable in the current microgravity
experiments \cite{lundblad,carollo2021}, as they are the cleanest platform to analyze such a BKT
finite-size scaling.
In particular, for the experimentally-relevant parameters in which the finite-size effects are sizable, we calculate the frequencies of the hydrodynamic excitations of the superfluid, that represent the main experimental probe of BKT physics \cite{hadzibabic,ozawa}. 
All these results are obtained with new and advanced techniques for the description of curved Bose-Einstein condensates, and contribute to the fundamental understanding of BKT and vortex physics, a paradigmatic topic in statistical physics, condensed matter physics and biology  \cite{kosterlitz2016,nelson:book,tan}. 

\section{Microgravity bubble-trapped condensates}
\begin{figure}[t]
\includegraphics[scale=0.2]{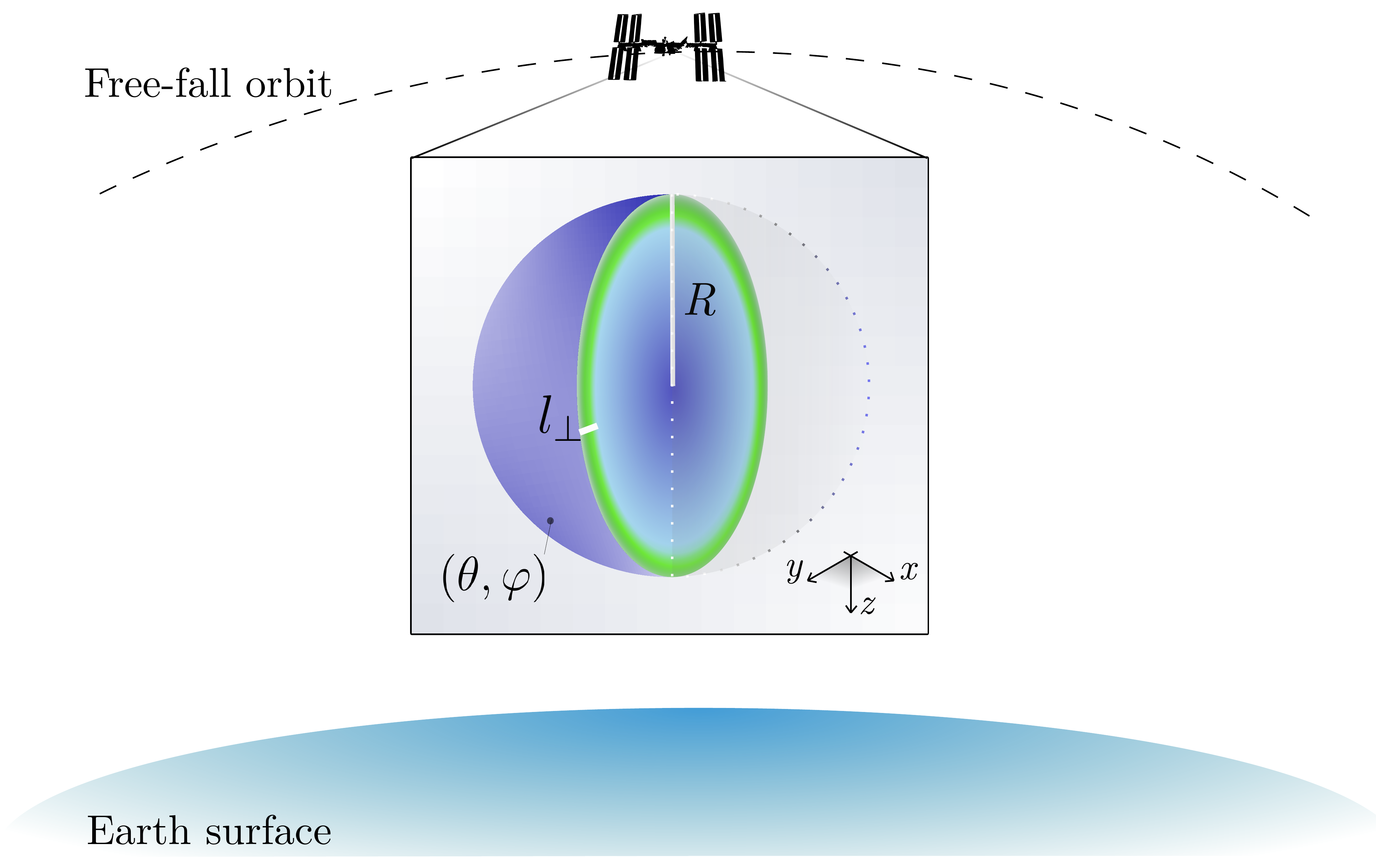}
\caption{Bubble-trapped Bose-Einstein condensates can be experimentally realized in microgravity conditions, where the gravitational contribution to the external potential is negligible.  
In this work, we model a two-dimensional spherical shell with radius $R$, thickness $l_{\perp} \ll R$, and we employ a system of spherical coordinates $(\theta,\varphi) \in [0,\pi] \times [0,2\pi]$. 
}
\label{fig1}
\end{figure}

Our theoretical analysis of the topological BKT transition and of its observable consequences is motivated by the possibility of producing two-dimensional bubble-trapped Bose-Einstein condensates {\cite{carollo2021}}. 
These systems are the first realization of a compact and curved Bose gas, whose local curvature can be controlled, even in the spherical case, by tuning the mean radius of the shell. In the absence of the gravitational contribution to the external potential, made possible by the microgravity conditions of the experiments \cite{carollo2021,lundblad}, the atoms can be confined on a spherically-symmetric shell with a bubble-trap potential \cite{zobay} $U(\vec{r}) = \sqrt{(m \omega_r^2 r^2/2-\hbar \Delta)^2+\hbar^2 \Omega_r ^2}$,
where $\Delta$, $\Omega_r $ and $\omega_r$ are experimentally tunable frequencies \cite{lundblad}, $m$ is the atomic mass, and $\hbar$ is Planck constant. 
In the limit of a thin large shell, $\Delta > \Omega_r $, the external potential coincides with the radially-shifted harmonic trap \cite{tononi2,sun}
$U_{\text{thin}} (\vec{r}) = m \omega_{\perp}^2 (r-R)^2 /2$,
where $\omega_{\perp} = \omega_r (2 \Delta/\Omega_r )^{1/2}$, {$R = [2\hbar\Delta/(m \omega_r^2)]^{1/2}$} is the radius of the spherical shell, and $l_{\perp} = [\hbar/(m\omega_{\perp})]^{1/2}$ is {its always finite thickness, see also the early study in Ref.~\cite{mitra2008}.}
To visually illustrate these systems we show a schematic setup in Fig.~\ref{fig1}. 

The results of the following sections rely on the derivation of the beyond-mean-field equation of state, which, to avoid introducing many technical details, is reported in the Appendix \ref{appendixA}. 
With this background, we will discuss the finite-size BKT transition of condensate shells, analyzing universal results that, by definition, do not depend on the microscopic details of the bosonic system.  
Nonetheless, to feature the typical parameters of the setup and to facilitate the comparison with the possible regimes of the experiments, we will also present nonuniversal results, obtained for specific values of the trap frequencies.
In this case, unless differently specified, we will consider $^{87}$Rb atoms and we will fix {$\omega_{r} = 2\pi \times 173 \, \text{Hz}$}, {$\Delta = 2\pi \times 30 \, \text{kHz}$}, {$\Omega_r = 2\pi \times 3 \, \text{kHz}$} {\cite{lundblad}, so that $R=15 \, \mu \text{m}$ and $l_{\perp} = 0.4 \, \mu \text{m}$. }

\section{Universal scaling of finite-size BKT physics}
In two-dimensional atomic superfluids at finite temperature, the possibility of establishing a superfluid flow relies on the absence of a turbulent vorticity field: in its presence, indeed, the ordering associated to a coherent flow of the quantum liquid is unattainable. 
The analysis of Kosterlitz and Thouless \cite{kosterlitz1973,kosterlitz1974}, that we extend here to the spherically-symmetric shell-shaped case, is devoted to understanding the statistical properties of vortices, whose appearance is regulated by the system temperature. 

Creating a vortex-antivortex pair at a distance equal to their core size $\xi$ requires an energy $2\mu_v$ and, in the low-temperature regime of $\beta^{-1}=k_B T \ll \mu_v$, where $T$ is the temperature and $k_{\text{B}}$ the Boltzmann constant, the vortex fugacity $y_0 = e^{-\beta \mu_v}$ is a small parameter. 
As a consequence, a superfluid has practically zero vorticity at low temperature and the superfluid density is well approximated by the \textit{bare} one, which is denoted by $n_s^{(0)}$, and is {given by a microscopic derivation of} the Landau formula {\cite{tononi1,tononi2018}}. 
As the temperature is increased, however, it becomes easier to excite vortices, whose physics is equivalent to that of a classical Coulomb gas on a sphere. 

Correspondingly, the bare fugacity $y_0$ is renormalized to $y(\theta)$, {with $\theta \in [0,\pi]$ the spherical coordinate, while} the bare superfluid density, expressed through the adimensional parameter $K_0 = \hbar^2 n_s^{(0)}/(m k_B T)$, is screened by a factor equal to the superfluid dielectric constant $\epsilon(\theta)$: $K_0$ becomes $K(\theta) = K_0 /\epsilon(\theta)$. 
The dependence on $\theta$ signals the running of the bare parameters $y_0$ and $K_0$ due to the existence of vortex-antivortex dipoles in the superfluid with distance $\theta^{'}<\theta$. 

We outline in Appendix \ref{appendixB} the derivation of the renormalization group equations that describe how $K(\theta)$ and $y(\theta)$ change with the system scale. We find
\begin{equation}
\begin{aligned}
\frac{\partial K^{-1}(\theta)}{\partial {\ell(\theta)}} &= 4 \pi^3 y^2 (\theta),
\label{Kfinal}
\\
\frac{\partial y(\theta)}{\partial {\ell(\theta)}} &= [2 - \pi K(\theta)] \, y(\theta),
\end{aligned}
\end{equation}
that extend the Kosterlitz-Nelson equations \cite{nelson} to a spherical superfluid. 
We introduce here the renormalization group scale $\ell(\theta)= \ln[(2R/\xi)\sin(\theta/2)]$, which depends on the 3D vortex-antivortex distance rather than on the geodesic distance $\theta$ \cite{nelsonvitelli,bereta}, a condition which is made possible only by the curvature of the superfluid. 
Let us now discuss the finite-size BKT transition, highlighting the role of the spherical geometry. 

\begin{figure}[hbtp]
\includegraphics[scale=1.0]{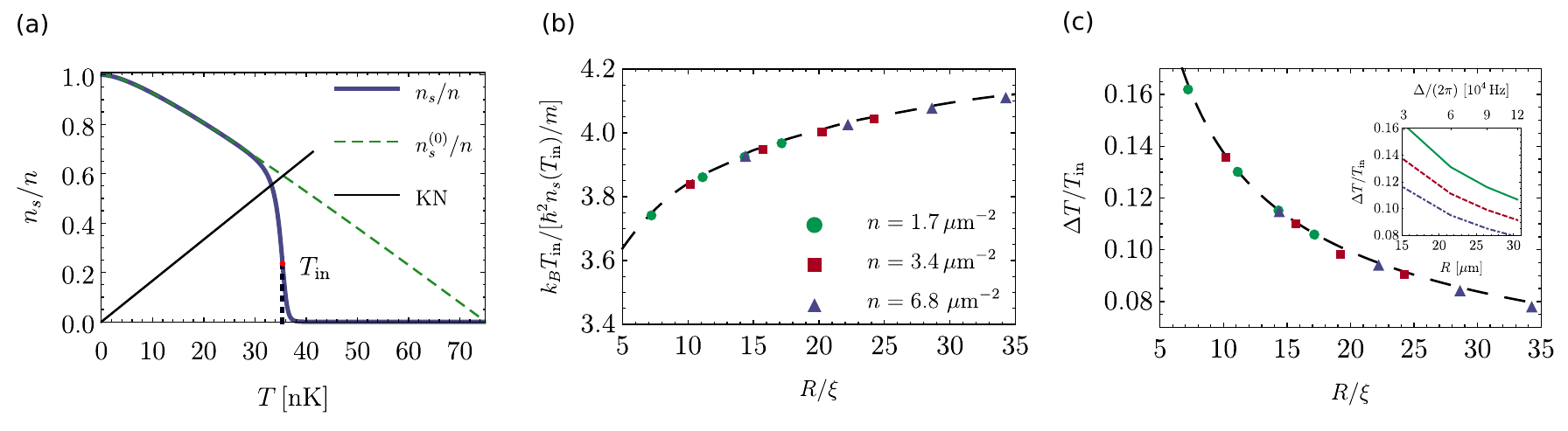}
\caption{Topological superfluid transition of a thin bubble-trapped superfluid. 
(a) The renormalized superfluid fraction $n_s/n$ of a shell-shaped superfluid does not display an abrupt jump at the Kosterlitz-Nelson temperature (KN) where $K(\pi) = 2/\pi$, but vanishes smoothly around the temperature $T_{\text{in}}$ of the inflection point. 
(b) {For different densities, shell widths, and rescaled radii $R/\xi$, implying different effective interactions $g_{\text{2D}}$, the values of $k_{\text{B}} T_{\text{in}}/[\hbar^2 n_s(T_{\text{in}})/m]$ collapse on the same curve {(the dashed line is a guide to the eye)}}. 
(c) The relative width of the superfluid transition $\Delta T/T_{\text{in}}$, 
goes to zero in the thermodynamic limit of $N,R \to \infty$, with $n$ fixed, 
{ and all} 
the points collapse on the interpolating dashed curve {$\Delta T /T_{\text{in}} = 2.1/\ln^2(5 R/\xi)$} if the abscissa is rescaled as $R/\xi$ {\cite{szeto1985,bramwell1994,komura2012}}, while the inset shows the explicit dependence on $R$ and on $\Delta$. 
We use in (a) the parameters of the first section and, for {$N=2 \times 10^4$ atoms, we find $T_{\text{in}} = 35 \, \text{nK}$, and $\Delta T = 4 \, \text{nK}$.} }
\label{fig2}
\end{figure}

We solve numerically Eqs.~\eqref{Kfinal} in the interval {$[\ell(\xi/R),\ell(\pi)]$}, using the bare superfluid density $n_s^{(0)}$ as the initial condition at {$\ell(\xi/R) \simeq 0$}, and obtaining the renormalized superfluid density $n_s (T)= m k_B T K(\pi)/\hbar^2$. 
As a consequence of the finite system size, {$\ell(\theta)$ can run only up to the finite scale $\ell(\pi)$, and $n_s(T)$ vanishes smoothly in a finite temperature region. 
This behavior is shown in Fig.~\ref{fig2}(a), where the inflection point of the superfluid fraction, i.~e.~$T_{\text{in}}$, estimates the location of the transition \cite{foster2010}, and where the finite transition width $\Delta T$ corresponds to the intersection between $n_s/n =0$, $n_s/n=1$ and the tangent in $T_{\text{in}}$.}
Although one would expect the broadening of the superfluid transition to be nonuniversal, i.~e.~system dependent, a proper rescaling of the system variables highlights the collapse of the data for different densities $n$ and for different ratios of $R/\xi$. 

For the following discussion, we need an explicit model of the vortex core size $\xi$. We identify it with the healing length 
$\xi = [\hbar^2/(2m g_{\text{2D}}n)]^{1/2}$, and, considering relatively large shells, we analyze the regime of $R \gg \xi$, in which the scattering properties of the bosonic gas can be approximately described as those of the flat 2D system. Thus, we adopt the mean-field interaction strength of flat 2D superfluids, which reads $g_{\text{2D}} = - 4\pi \hbar^2/[m \ln(n a^2)]$ \cite{pitaevskii}, with $a$ the $s$-wave scattering length on the shell \cite{zhang}, assumed $
a = 2 (\pi/C)^{1/2} \exp[-\gamma-(\pi/2)^{1/2} \,  l_{\perp}/a_{\text{3D}}] \, l_{\perp}$,
where {$a_{\text{3D}} = 5 \, \text{nm}$, and} $C=0.915$ obtained by solving the two-body scattering problem \cite{petrov}. 

As we show in Fig.~\ref{fig2}(b), the ratio $m k_{\text{B}} T_{\text{in}}/[\hbar^2 n_s(T_{\text{in}})]$ scales in an universal way with $R/\xi$, revealing that $n_s(T_{\text{in}}) \to n_s(T_{\text{BKT}}^{+}) = 0$ at the thermodynamic limit, 
with $T_{\text{BKT}}$ the infinite-system transition temperature. 
Moreover, as Fig.~\ref{fig2}(c) depicts, also the relative width of the transition region $\Delta T /T_{\text{in}}$ scales logarithmically with $R/\xi$ as $\Delta T /T_{\text{in}} = c_1/\ln^2(c_2 R/\xi)$, with $c_1$ and $c_2$ adimensional parameters \cite{szeto1985,bramwell1994,komura2012}. 
The previous relation can be derived from the correlation length $\xi_c$ of the infinite superfluid \cite{kosterlitz1974}, i.~e.~$\xi_c/R= c_2  \exp[(c_1/\tau)^{1/2}]$, by assuming that $\tau=\Delta T/T_{\text{in}}$ when $\xi_c = \xi$. 

In our experimentally-relevant regimes, the scalings of Figs.~\ref{fig2}(b), \ref{fig2}(c) are universal, i.~e.~the collapse appears also if $l_{\perp}$ and the interaction strength $g_{\text{2D}}$ are changed, but $c_{1,2}$ depend in a nontrivial way on 
$\omega_{r}$ and $\Omega_r$, that we suppose fixed \cite{lundblad}. 
The independence on $g_{\text{2D}}$ and on $l_{\perp}$ suggests that our findings are qualitatively relevant for ellipsoidal shells \cite{carollo2021}. 
We have actually found a similar logarithmic scaling in box-trapped superfluids  \cite{furutani2021}, which will be analyzed in detail in Ref.~\cite{tononi}. 
Concerning other geometries, a BKT transition should not necessarily occur in cylindrical or in large toroidal superfluids \cite{machta1989}. 

\onecolumngrid

\twocolumngrid

These universal laws are of immediate experimental interest, as they are obtained with the first derivation of the BKT Eqs.~\eqref{Kfinal} for shell-shaped condensates. 
Our whole analysis relies on a {microscopic derivation of the corresponding beyond-mean-field equation of state} and on the extension of BKT theory to the spherical case (see Appendices \ref{appendixA} and \ref{appendixB}). 
We emphasize that our purely two-dimensional description is valid if 
{$\{g_{\text{2D}} \, n, k_{\text{B}} T_{\text{in}} \} < \hbar \omega_{\perp}$} \cite{pitaevskii}, and that the scattering properties are described correctly for $R \gg \xi$. 
The previous inequalities, considering for instance the values of $l_{\perp}$ and of $R$ of the previous section, imply that the number of particles must be in the interval {$2 \times 10^2 \ll N \lesssim 2  \times 10^4$}. 
In all our simulations, in which we change the number density $n$, we choose the appropriate number of atoms to ensure that the system is two-dimensional and that the condition $R \gg \xi$ is satisfied. 

\section{Probing the superfluid BKT transition}
The occurrence of the superfluid BKT transition and the universal scaling of the finite-size corrections in bubble-trapped superfluids can be experimentally probed with qualitative and quantitative means. 
We expect that, in the temperature-broadened BKT transition region, the absorption imaging of the expanding shell will produce a wavy interference pattern \cite{hadzibabic2}, different from the zero-temperature one \cite{tononi2}, which will reflect the specific configuration of the vortices. This qualitative observation will prove that the vortex-antivortex unbinding is the driving mechanism of the superfluid transition in two-dimensional compact shells. 
A more quantitative description of the superfluid density and of its vanishing at the BKT transition can be achieved by measuring the temperature dependence of the first and second sound \cite{hadzibabic,ozawa}.
To investigate this phenomenon in the context of bubble-trapped superfluids, we need first to extend the Landau two-fluid model \cite{landau1941}.

\subsection{Two-fluid model description of a superfluid shell}
Let us extend Landau two-fluid model \cite{landau1941} to a bubble-trapped superfluid, with the aim of calculating the frequencies of the hydrodynamic modes $\omega$. 
We assume that, after a small perturbation of the superfluid shell, the system remains locally at equilibrium, so that the thermodynamic variables acquire an additional dependence on the time $t$ and on the spherical coordinates ($\theta$,$\varphi$). Following Landau, Ref.~\cite{landau1941}, we obtain two wave equations for the superfluid shell: 
$\partial^2 \rho/\partial t^2 = (-\hat{L}^2 P) /(R^2\hbar^2) $, 
that describes the dynamics of pressure-density oscillations, with $\rho$ being the mass density and $P$ the pressure, and 
$\partial^2 \tilde{s}/\partial t^2 = (n_s/n_n) \tilde{s}^2 \, (-\hat{L}^2 T) /(R^2\hbar^2)$,
describing the evolution of temperature-entropy oscillations, with $\tilde{s}$ the entropy per unit of mass and atom number and $n_n = n - n_s$ the renormalized normal density. Note that these differential equations contain the angular momentum operator in spherical coordinates, i.~e.~$\hat{L}^2$. 

We now linearize these equations by decomposing all the thermodynamic variables as $x(\theta,\phi,t) = x_0 + x'(\theta,\phi,t)$, with $x_0$ the equilibrium value, and $x' \ll x_0$ the small elongation field. Then, we rewrite the wave equations by expressing the fluctuations of the thermodynamics variables in terms of $P'$ and $T'$ only, obtaining a system of two differential equations in $P'$ and in $T'$. 
The dynamics of an excited superfluid shell can be decomposed as the superposition of orthonormal modes in the basis of spherical harmonics $Y_l^{m_l} (\theta,\phi)$: rewriting the elongation fields as $P^{'} = \tilde{P}(\omega,l,m_l) \, e^{i \omega t} \, Y_l^{m_l} (\theta,\phi)$ and $T^{'} = \tilde{T}(\omega,l,m_l) \, e^{i \omega t} \, Y_l^{m_l} (\theta,\phi)$, the system of differential equations becomes algebraic. Setting the system determinant to zero, we obtain the following biquadratic equation
\begin{equation}
{
\omega^4 - \omega^2 \, \frac{( v_A^2 
+ v_L^2 )}{R^2} \, [l(l+1)]
+ \frac{v_A^2 v_L^2}{R^4} \, \frac{[l(l+1)]^2}{\kappa_T/\kappa_{\tilde{s}}} = 0, 
}
\label{eqsound}
\end{equation}
which extends the familiar result of Landau \cite{landau1941} to a spherically-symmetric superfluid. In particular, here we define the adiabatic velocity $v_A^2 = (\partial P/\partial \rho)_{\tilde{s}}$, {the isothermal compressibility $\kappa_T$ and the adiabatic one $\kappa_{\tilde{s}}$,} and the Landau velocity $v_L^2 = n_s T \tilde{s}^2 /(n_n \tilde{c}_v )$, with $\tilde{c}_v$ the specific heat per unit of mass and atom number. Note that all the thermodynamic quantities can be derived from the equation of state at finite temperature (see Appendix \ref{appendixA}).

By solving Eq.~\eqref{eqsound}, we obtain the frequencies of the hydrodynamic excitations of frequency $\omega_1$ and $\omega_2$ ($< \omega_1$), which read 
{
\begin{equation}
\omega_{1,2}^2 = \bigg[ \frac{l(l+1)}{R^2} \bigg] \bigg[ \frac{v_{A}^2 + v_{L}^2}{2} \pm \sqrt{\bigg(\frac{v_{A}^2 + v_{L}^2}
	{2}\bigg)^2 - \frac{v_{A}^2 v_L^2}{\kappa_T/\kappa_{\tilde{s}}}} \bigg].
\label{surfaceomegas}
\end{equation}
}
We emphasize that, in the spherical case, the dispersion relation between $\omega$ and $l$, with $l$ the quantum number of angular momentum, is not linear, and the familiar first and second sound modes correspond to hydrodynamic excitations of frequency $\omega_1$ and $\omega_2$.

\begin{figure}
\includegraphics[scale=1]{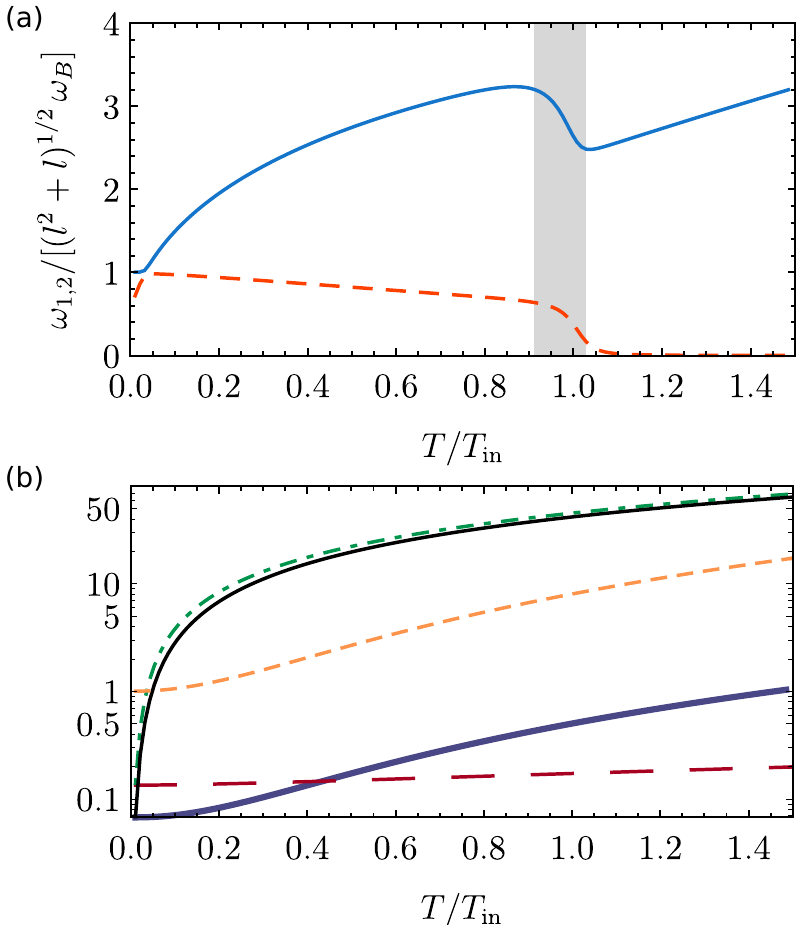}
\caption{Hydrodynamic surface modes at finite temperature. (a) Frequency of the first and second excited modes $\omega_{1,2}$ plotted as a function of $T$, and rescaled with $(l^2+l)^{1/2}\omega_B$, where $\omega_B = \sqrt{\mu/(mR^2)}$, and $\mu$ is the chemical potential. The shaded area highlights the region of width {$\Delta T/T_{\text{in}}$}, in which{, for the specific parameters of Fig.~\ref{fig2}(a), the superfluid density vanishes.} the superfluid density of Fig.~\ref{fig2}(a) vanishes. 
(b) While the surface modes are non-monotonic around the infinite-order BKT transition, the thermodynamic functions are monotonic: 
{
we plot the entropy $\tilde{s}$ per unit of mass and atom number (black thin), the specific heat $\tilde{c}_V$ per unit of mass and atom number (green dot-dashed), the ratio  of isothermal and adiabatic compressibilities $\kappa_T/\kappa_{\tilde{s}}$ (orange dashed), the chemical potential $\mu/E_{l_{\perp}}$ (red long-dashed), and the grand potential $-\Omega/E_{l_{\perp}}$ (blue thick).
}
The physical parameters chosen for this figure are the same of Fig.~\ref{fig2}(a), and we define $E_{l_{\perp}} = \hbar^2/(2ml_{\perp}^2)$.}
\label{fig3}
\end{figure}

\subsection{Hydrodynamic excitations of bubble-trapped superfluids}
By making use of the equation of state, whose derivation is outlined in Appendix \ref{appendixA}, and fixing the trap parameters of Fig.~\ref{fig2}(a), we explicitly calculate the frequencies $\omega_1$ and $\omega_2$. 
In Fig.~\ref{fig3}(a) we plot the hydrodynamic frequencies as a function of the temperature. While, as a consequence of the vanishing superfluid density, the surface modes are not monotonic around $T_{\text{in}}$, the thermodynamic functions are monotonic, see Fig.~\ref{fig3}(b). 
Note that, in the limit of a very large radius of the sphere, these frequencies would converge to the familiar first and second sound of flat superfluids \cite{pitaevskii}.
We also emphasize that these hydrodynamic excitations consist of a displacement of the atoms along a tangent direction to the shell itself: we assume that the radius $R$ remains fixed during the dynamics and that the experimental protocol does not excite any radial motion. 
Clearly, it is impossible to observe the higher-energy surface modes without violating this hypothesis, and only the surface modes with frequency $\omega \lesssim \omega_{\perp}$ can be observed. 
Following this criterion, we predict that in the ongoing experiments, whose typical trap parameters are outlined in Fig.~\ref{fig2}(a), it will be possible to observe up to the $l=3$ modes. 

\section{Conclusions}
The experiments with quantum gases, from NASA-JPL Cold Atom Lab \cite{aveline}, which allows for a clean production of single shells \cite{carollo2021,lundblad}, and, in perspective, from BECCAL \cite{frye}, are the ideal platform to explore the superfluid BKT transition in finite-size systems and to analyze its qualitative and quantitative consequences. 
This line of research is highly relevant, as the past experiments with superfluid helium adsorbed in porous materials, modeled as a collection a packed spheres, did not produce a unique conclusion on the existence and on the nature of finite-size BKT physics in shells \cite{bishop,kotsubo} {(see also the note \cite{ovrutnote})}. 
It is also important to stress that the study of these finite-temperature properties is not only relevant, but also necessary: in bubble-trapped condensates, the typical temperatures at which the regime of quantum degeneracy is reached are lower in comparison with the more familiar harmonically-trapped condensates \cite{tononi2,rhyno2021,carollo2021}. 
In conclusion, our work contributes to the fundamental understanding of the Berezinskii-Kosterlitz-Thouless transition in finite size systems and, specifically, in topologically-nontrivial bubble-trapped superfluids.
Their {analysis} in microgravity facilities \cite{carollo2021} will offer the definitive proof that, despite the peculiar topology, a thin two-dimensional shell can host a superfluid flow at finite temperature, and that the basic mechanism of the superfluid transition relies on the vortex-antivortex unbinding of the seminal BKT theory. 
The quantitative understanding of the superfluid properties in two-dimensional shell-shaped condensates is the first fundamental step in the development of analogue simulation of quantum many-body physics with curved quantum gases. 

\begin{acknowledgments}
The authors acknowledge useful discussions with A. Fetter, T.-L. Ho, N. Lundblad, and D. R. Nelson. 
A.~P. ~acknowledges financial support by the Deutsche
Forschungsgemeinschaft (DFG, German Research Foundation) via the
Collaborative Research Center SFB/TR185 (Project No. 277625399). A.T.
acknowledges the support of the ANR grant Droplets (19-
CE30-0003).
\end{acknowledgments}

\appendix

\section{Equation of state of a 2D spherical shell}
\label{appendixA}
We derive the equation of state of a shell-shaped Bose gas starting from the beyond-mean-field grand potential, $\Omega$, obtained at a one-loop level in Ref.~\cite{tononi1}.
Thus, we consider here $\Omega = \Omega_0 + \Omega_g^{(0)} + \Omega_g^{(T)}$, where $\Omega_0 = - 4\pi R^2 \mu^2/(2 g_0)$ is the mean-field contribution, with $\mu$ the chemical potential and $g_0$ the bare interaction strength. The beyond-mean-field contribution at zero temperature, including the counterterms produced by convergence factor regularization \cite{salasnich}, reads 
\begin{equation}
\Omega_g^{(0)} = \frac{1}{2}\sum_{l=1}^{\infty} \sum_{m_l=-l}^{l} \big( E_l -\epsilon_l -\mu \big),
\end{equation}
where $E_l^{B} = \sqrt{\epsilon_l(\epsilon_l+2\mu)}$ is the Bogoliubov spectrum, with $\epsilon_l = \hbar^2 l(l+1)/(2mR^2)$. The beyond-mean-field contribution at finite temperature is 
\begin{equation}
\Omega_g^{(T)}= \frac{1}{\beta} \sum_{l=1}^{\infty} \sum_{m_l=-l}^{l} \ln(1-e^{-\beta E_l^B}). 
\end{equation}

To find the renormalized grand potential $\Omega$ we repeat the scattering-theory \cite{landau} procedure outlined in Refs.~\cite{mora,mora2}. In particular, in the sum over $l$ of $\Omega_g^{(0)}$ we introduce the ultraviolet cutoff $l_c$, and we integrate instead of summing. 
In terms of this cutoff, we also obtain the bare interaction strength 
\begin{equation}
g_0 = - \frac{2\pi\hbar^2}{m} \frac{1}{{\ln[\sqrt{l_c(l_c+1)}\, a \, e^{\gamma}/(2R)] }},
\end{equation}
where $a$ is the two-dimensional $s$-wave scattering length on the sphere \cite{zhang}. 
The logarithmic divergence of $\Omega_g^{(0)} \sim \ln[l_c(l_c+1)]$ is perfectly balanced by $\Omega_0$, which is also divergent through its dependence on $g_0 \sim 1/\ln[l_c(l_c+1)]$. Therefore, the dependence on the ultraviolet cutoff $l_c$ cancels out, and we find the renormalized grand potential per unit of area: 
\begin{eqnarray}
&&\frac{\Omega}{4 \pi R^2}  = -\frac{m\mu^2}{8\pi\hbar^2} \bigg[ \ln \bigg( \frac{4\hbar^2}{m\mu a^2 \, e^{2\gamma+1}}\bigg) +\frac{1}{2}\bigg] 
\nonumber
\\
&+&\frac{m E_1^B}{8\pi\hbar^2} (E_1^B -\epsilon_1 - \mu) + \frac{m\mu^2}{8\pi\hbar^2} \ln \bigg( \frac{E_1^B +\epsilon_1 + \mu}{\mu}  \bigg)
\nonumber
\\
&+&\frac{1}{4\pi R^2}\frac{1}{\beta} \sum_{l=1}^{\infty} \sum_{m_l=-l}^{l} \, \ln (1 - e^{-\beta E_l^B}).
\label{omegasphere}
\end{eqnarray}
In the limit of infinite radius, Eq.~\eqref{omegasphere} coincides with the flat-case result of Ref.~\cite{mora2} at a one-loop level.

From the grand potential $\Omega(\mu,V,T)$, all the thermodynamic quantities can be obtained with standard thermodynamic relations. 
In particular, for a fixed area $V=4 \pi R^2$ and temperature $T$, we calculate the number of particles $N=- (\partial \Omega/\partial \mu)_{V,T}$ as a function of the chemical potential $\mu$. Inverting numerically this relation, we obtain the chemical potential $\mu(N,V,T)$, which allows us to derive the Bogoliubov spectrum $E_l^{B}$ and to obtain the free energy with a standard Legendre transformation
$F(N,V,T) = \mu(N,V,T) N + \Omega (\mu(N,V,T),V,T)$. 
Consequently, the entropy per unit of mass and atom number reads $\tilde{s} = - (m N)^{-1} (\partial F/\partial T)_{N,V}$, the specific heat per unit of mass and atom number is $\tilde{c}_v = - T/(m N) ( \partial^2 F/\partial T^2)_{N,V}$, and the pressure reads 
$P = - ( \partial F/\partial V)_{N,T}$.

\section{Vortices and BKT transition}
\label{appendixB}
We introduce here the bare superfluid density and the vortex chemical potential, and we discuss how to derive the renormalization group equations that describe how these bare parameters are renormalized by the topological excitations of the superfluid, i~e.~the vortices. 

\subsection{Bare superfluid density and vortex chemical potential}
\label{appendixB1}
Let us derive the bare superfluid density $n_s^{(0)}$ and the vortex chemical potential $\mu_v$ of a bubble-trapped superfluid. For this scope, we need to model the additional kinetic energy contribution $E^{\text{(vor)}}$ of a system of vortices in a spherical superfluid, a contribution which is neglected in $\Omega$, that takes into account only the Bogoliubov quasiparticles. We calculate $E^{\text{(vor)}}$ as
\begin{equation}
E^{\text{(vor)}} = \frac{1}{2} m n_s^{(0)} \int_0^{2\pi} d\varphi \int_0^\pi d\theta  \, R^2  \sin \theta  \, (\vec{v} \cdot \vec{v}), 
\label{evor}
\end{equation} 
where the bare superfluid density $n_s^{(0)}$ is a low-temperature approximation of the renormalized one, $n_s$, and reads \cite{tononi1}
\begin{equation}
n_s^{(0)} = n - \beta \sum_{l=1}^{\infty} \, \frac{(2l+1)}{4\pi R^2} \, \frac{\hbar^2 l (l+1)}{2 m R^2} \, \frac{e^{\beta E^{B}_l }}{(e^{\beta E^{B}_l }-1)^2}, 
\label{landauns}
\end{equation}
which is derived with a microscopic quantum-field-theory calculation analogous to that of Ref.~\cite{tononi2018}.
To obtain the chemical potential of the vortices $\mu_v$, we need to integrate Eq.~\eqref{evor} and, therefore, we have to derive the velocity field $\vec{v}$ of the superfluid part of the fluid with nonzero vorticity.
We model the velocity of an incompressible vorticous superfluid as   
$\vec{v} = 2 \pi \hbar/m \, \hat{r} \times \vec{\nabla}_{\text{R}} \, \chi (\theta,\varphi)$, where $\vec{\nabla}_{\text{R}}$ is the gradient in spherical coordinates \cite{moller} and $\chi (\theta,\varphi)$ is the stream function \cite{bereta}, constant along the streamlines of the fluid. 
For a system of $M$ vortices with zero global charge, with vortex cores at $(\theta_1,\varphi_1), ... ,(\theta_M,\varphi_M)$ and charges $q_1,...,q_M$, the stream function reads $\chi(\theta,\varphi) = \sum_{i=1}^{M} \chi_i(\theta,\varphi)$
with $\chi_i(\theta,\varphi) = q_i /(2\pi) \, \ln [ \sin (\gamma_i/2)]$, 
and $\gamma_i$ is the angular distance between $(\theta,\varphi)$ and $(\theta_i,\varphi_i)$ \cite{bereta}. 
Given the velocity field $\vec{v}$ associated to a configuration of vortices with stream function $\chi(\theta,\varphi)$, we calculate the kinetic energy of the vortical configurations $E^{\text{(vor)}}$ \cite{nelsonvitelli,bereta}, obtaining
\begin{equation}
\begin{aligned}
E^{\text{(vor)}} = &\sum_{i=1}^{M} q_i^2 \, \mu_v  
\\
- \frac{\hbar^2 \pi n_s^{(0)}}{m} &\sum_{ \substack{i,j=1 \\ i \neq j}}^{M}  q_i q_j \,  \ln \bigg[ \frac{2R \, \sin (\gamma_{ij}/2)}{\xi}  \bigg] ,
\label{EwithxiR}
\end{aligned}
\end{equation}
where, due to $R \gg \xi$, we deduce the flat-case value of the vortex chemical potential $\mu_v = \hbar^2 n_s^{(0)} \, \pi \, [\ln (2 \sqrt{2}) + \gamma]/m $ \cite{kosterlitz1973}, with $\gamma$ the Euler-Mascheroni constant. 
Note that $2\mu_v$ is the energy required to create a couple of vortices with charges $q_1=-q_2 = 1$ at the minimal distance. 

\subsection{Renormalization of the bare parameters}
\label{appendixB2}
Following the works of Kosterlitz and Thouless \cite{kosterlitz1973,kosterlitz1974}, we now discuss the main technical steps to derive the renormalization group equations of the main text.

The fundamental correspondence between a superfluid with nonzero vorticity and a classical Coulomb gas, on which the analysis of Kosterlitz and Thouless is based, holds also in shell-shaped systems. 
Thus, we reinterpret the kinetic energy $E^{\text{(vor)}}$ of the vorticous superfluid as the interaction energy between charged particles on a shell. 
In particular, we focus on a vortex-antivortex dipole on the surface of the sphere with charges $q_1=-q_2=1$, whose cores are at the positions $(\theta_1,\varphi_1)=(\theta,0)$ and $(\theta_2,\varphi_2)=(0,0)$, and whose bare interaction energy is
$\beta U_0(\theta) = 2 \beta \mu_v + 2\pi K_0 \,  \ln [ 2R \, \sin (\theta/2)/\xi]$, with $K_0$ defined in the main text. 
Due to the polarization of the medium at $\theta' < \theta$, the force between the dipoles is screened as $d U (\theta)/d \theta = \varepsilon(\theta)^{-1} dU_0 (\theta)/d\theta$, where $\varepsilon(\theta)$ is the relative dielectric constant, and \cite{young,kotsubo}
\begin{equation}
\beta U(\theta) = 2\pi  \int_{{\ell(\xi/R)}}^{{\ell(\theta)}} K(\theta') \ \text{d}[{\ell(\theta')}],
\label{renoU}
\end{equation}
with $K(\theta)$ defined in the main text,
and where the logarithm of the distance between the vortices, $\ell(\theta)=\ln[(2R/\xi)\sin(\theta/2)]$, appears explicitly as the new integration variable. 

In two-dimensional electrostatics the dielectric constant $\varepsilon(\theta)$ is related to the susceptibility $\chi(\theta)$ as $\varepsilon(\theta) = 1 + 4\pi \chi(\theta)$, where $\chi(\theta) = \int_{\xi/R}^{\theta} d\theta'  n_d(\theta') \, \alpha(\theta')$ is obtained integrating the product of $n_d(\theta')$, the density of dipoles with angular separation $\theta'$, and of $\alpha(\theta')$, the polarizability of the medium at separation $\theta'$. 
Following Ref.~\cite{kotsubo}, we find
$n_d(\theta') =  2 \pi  (R^2/\xi^4) \sin \theta' \, y_0^2 \, e^{-\beta U(\theta')}   + o(y_0^4)$.
But, interpreting the dipole moment differently from Ref.~\cite{kotsubo}, as the product between the vortex charge and the 3D distance among the charges, we get $\alpha (\theta') = 2\pi K_0 \, R^2 \sin^2(\theta'/2)$. 
Taking into account this crucial difference we derive $\chi(\theta)$, inserting it in $\varepsilon(\theta)$ and finally, at the lowest order in $y_0$, we find the renormalized $K(\theta)$:
\begin{equation}
K^{-1}(\theta) = K_0^{-1} + 4\pi^3  \int_{{\ell(\xi/R)}}^{{\ell(\theta)}} y^2(\theta') \ \text{d}[{\ell(\theta')}],
\label{KrenoM}
\end{equation}
where we define the square of the renormalized fugacity $y^2(\theta)$ as
\begin{equation}
y^2(\theta) = y_0^2  \ \frac{\sin^4(\theta/2)}{{[\xi/(2R)]^4}} \,  e^{-\beta U(\theta)}.
\label{yrenoM} 
\end{equation}
Note that Eqs.~\eqref{KrenoM}, \eqref{yrenoM} are analogous to those of Ref.~\cite{young}, derived in the flat case, and the renormalization group equations of the main text are obtained by differentiating Eqs.~\eqref{KrenoM}, \eqref{yrenoM} with respect to the distance between the vortices, i.e. $\ln[\sin(\theta/2)]$. 

To obtain the renormalized superfluid density $n_s$ we calculate the bare superfluid density $n_s^{(0)} (T)$, which fixes the values of the initial conditions $K_0$ and $y_0$. Thereafter, we solve numerically the renormalization group equations for each temperature $T$, obtaining the renormalized $K(\theta)$, and, from that, $n_s$.

\end{document}